\documentclass[%
aip,
amsmath,amssymb,
reprint,%
]{revtex4-1}

\usepackage{graphicx}
\usepackage{dcolumn}
\usepackage{bm}

\usepackage[utf8]{inputenc}
\usepackage[T1]{fontenc}
\usepackage{mathptmx}
\usepackage{etoolbox}

\usepackage[T1]{fontenc}
\usepackage[utf8]{inputenc}
\usepackage{graphicx}
\usepackage{xcolor}
\usepackage{hyperref}
\hypersetup{
colorlinks=true,   linkcolor=blue,   citecolor=blue,   urlcolor=blue }

\newcommand{\figref}[1]{FIG.~\ref{#1}}

\makeatletter
\def\@email#1#2{%
 \endgroup
\patchcmd{\titleblock@produce}   {\frontmatter@RRAPformat}   {\frontmatter@RRAPformat{\produce@RRAP{*#1\href{mailto:#2}{#2}}}\frontmatter@RRAPformat}   {}{} }%
\makeatother
\begin{document}

\preprint{AIP/123-QED}

\title{
Osprey: Production-Ready Agentic AI for Safety-Critical Control Systems }

\author{Thorsten Hellert}
\email{thellert@lbl.gov}

\affiliation{Lawrence Berkeley National Laboratory}

\author{João Montenegro}
\affiliation{Lawrence Berkeley National Laboratory}

\author{Antonin Sulc}
\affiliation{Lawrence Berkeley National Laboratory}

\date{\today}

\begin{abstract}
Operating large-scale scientific facilities requires coordinating diverse subsystems, translating operator intent into precise hardware actions, and maintaining strict safety oversight.
Language model–driven agents offer a natural interface for these tasks, but most existing approaches are not yet  reliable or safe enough for production use.
In this paper, we introduce Osprey, a framework for using agentic AI in large, safety-critical facility operations.
Osprey is built around the needs of control rooms and addresses these challenges in four ways.
First, it uses a plan-first orchestrator that generates complete execution plans, including all dependencies, for human review before any hardware is touched.
Second, a coordination layer manages complex data flows, keeps data types consistent, and automatically downsamples large datasets when needed.
Third, a classifier dynamically selects only the tools required for a given task, keeping prompts compact as facilities add capabilities.
Fourth, connector abstractions and deployment patterns work across different control systems and are ready for day-to-day use.
We demonstrate the framework through two case studies: a control-assistant tutorial showing semantic channel mapping and historical data integration, and a production deployment at the Advanced Light Source, where Osprey manages real-time operations across hundreds of thousands of control channels.
These results establish Osprey as a production-ready framework for deploying agentic AI in complex, safety-critical environments.
\end{abstract}

\maketitle

\begin{figure}[b!]
    \centering
    \includegraphics[width=\linewidth]{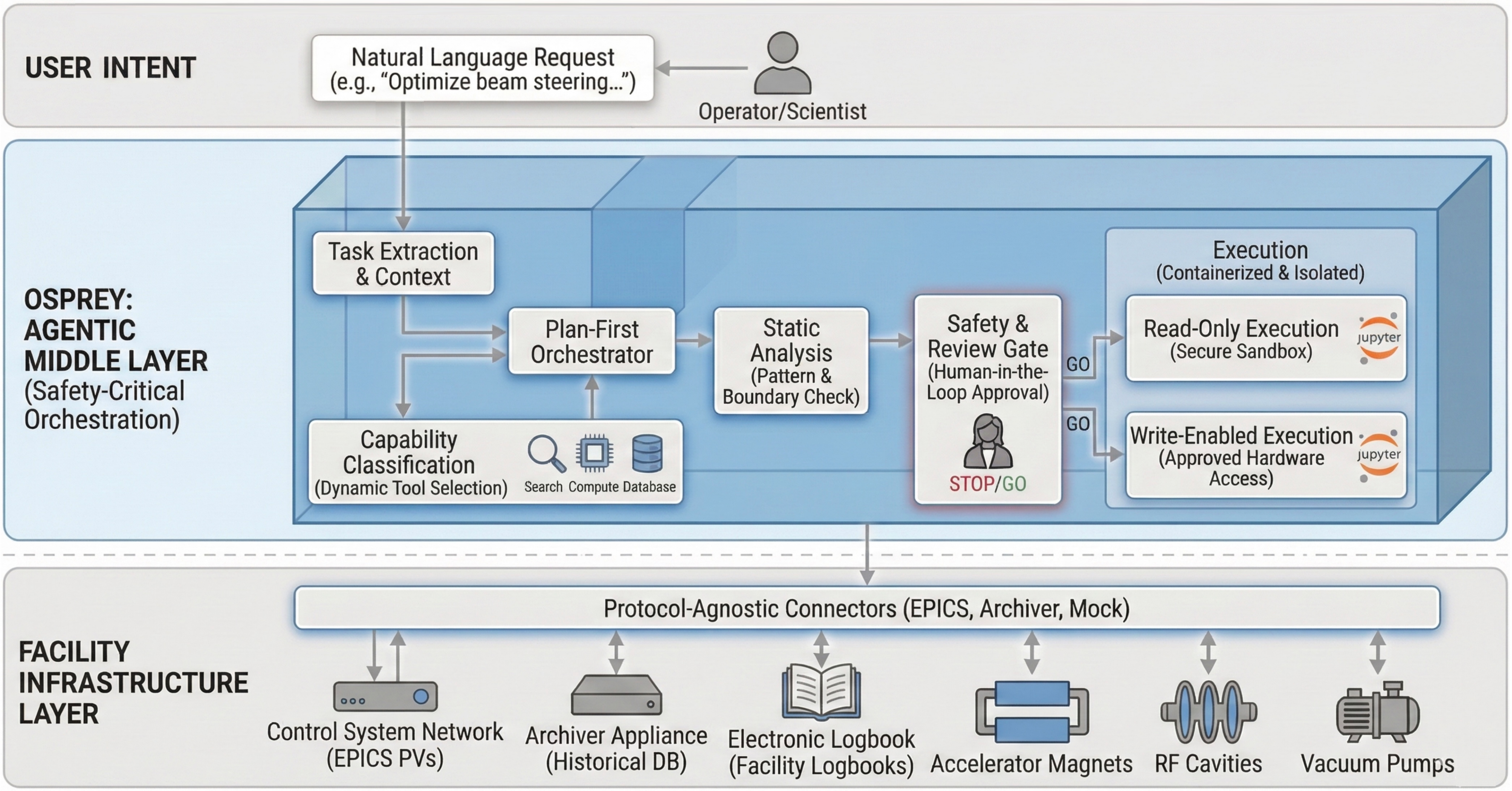}
    \caption{Osprey provides agentic orchestration with human-in-the-loop safety review, translating natural language requests into approved, isolated execution on facility control systems.}
    \label{fig:osprey}
\end{figure}

\section{Introduction}
\label{sec:intro}

Large scientific facilities such as particle accelerators~\cite{Nicklaus2024,Ostroumov2023}, fusion experiments~\cite{DEVRIES2024114464}, or large telescopes~\cite{ELT_controls} operate extremely complex technical infrastructures.
These environments expose tens of thousands to millions of control system channels (commonly called process variables or PVs in EPICS~\cite{EPICS}-based systems) across many subsystems, including magnets, RF systems, diagnostics, vacuum and cryogenics hardware, and safety interlocks.
Operators must translate high-level goals into precise sequences of actions on this underlying equipment.
Much of this knowledge resides in expert practice rather than formalized software, making operations labor-intensive and difficult to scale, particularly during fault recovery~\cite{Branlard2018RFOpXFEL} or other time-critical procedures~\cite{s25061666}.

Many of these facilities also carry inherent hazards.
Synchrotron-radiation storage rings and high-power accelerators store substantial beam energy, where uncontrolled beam loss or mis-steering can produce ionizing radiation or damage sensitive components.
Modern radiation-protection and machine-protection systems~\cite{Andersson2015MPSAvailability,Schmidt2017MachineProtection}, therefore, play a central role in maintaining operational safety and facility availability~\cite{Bozzato2024_RP_LHC, Cimmino2024_RP_Petawatt, Li2024_RealTimeBeamProtection}.
These environments are also subject to stringent regulatory oversight, meaning that any automated system interacting with hardware must operate within well-defined safety boundaries and undergo careful review before deployment.

Modern language models (LMs) and agentic AI systems~\cite{sapkota2025ai,yao2023react,wu2023autogen} offer intuitive natural-language interfaces and broad, cross-domain expertise that can exceed the situational overview of any single operator~\cite{Sulc2024xou}.
However, existing general-purpose agent frameworks often lack key properties needed for facility operations: they offer little visibility into planned future actions, lack protocol-aware access to control stacks such as EPICS~\cite{EPICS}, and provide only limited safeguards for hardware writes.
These constraints motivate agentic architectures that are not only language-driven but also explicitly aware of operational context.
To be viable in safety-critical facilities, such systems must provide transparent multi-step planning~\cite{rawat2025preact,langgraph2025}, work reliably with heterogeneous data sources and devices, enforce clear safety limits on all hardware operations, and scale to large capability inventories without prompt inflation~\cite{gan2025ragmcp} while propagating facility-resolved data through complex workflows and supporting both simple and advanced code-generation strategies~\cite{anthropic2025_claude_agent_sdk}.
Although our work is motivated by accelerator and experimental facilities, the underlying requirements are representative of many domains where complex technical systems must be steered safely and transparently.

To address these requirements, we introduce \emph{Osprey}~\cite{osprey_repo}, a framework designed for deploying agentic AI in large-scale, safety-critical facility operations (\figref{fig:osprey}).
Osprey integrates four core architectural components: (i) a plan-first orchestrator that produces explicit, reviewable execution plans before any action occurs; (ii) a capability classification layer that scales tool use through per-capability relevance decisions; (iii) connector abstractions that provide protocol-agnostic access to production control infrastructures, archiver and historical data acquisition systems, and mock environments; and (iv) a safe, containerized execution layer that isolates generated analysis code and enforces mandatory approval for hardware writes.
The framework further includes an extensible code-generation architecture and supports ecosystem integration through automated capability creation from Model Context Protocol (MCP) servers~\cite{mcp}.

We demonstrate Osprey through a deployable control assistant template with production-ready integration patterns~\cite{osprey_tutorial_channel_finder}, already adopted in early deployments at multiple facilities.
We also report on an operational deployment at the Advanced Light Source (ALS)~\cite{hellert2025agenticai}, where the framework supports real-time accelerator operations.
Together, these results show that agentic workflows can be integrated safely and reliably into demanding operational environments while remaining accessible to new facilities.


\section{Background and Related Work}
\label{sec:background}

\subsection{Agentic Systems and Tool Use}

The foundations of agentic language models and multi-agent systems build on the classic definition of agents as entities that perceive their environment and act upon it, with rational agents aiming to maximize performance measures~\cite{russelnorvig2025ai}.
Extending this principle, multi-agent systems introduce shared environments in which agents may cooperate or compete.
Early frameworks such as ReAct unified reasoning and acting in iterative loops for tool-based tasks~\cite{yao2023react}, as AutoGen~\cite{wu2023autogen} extended this paradigm toward multi-agent conversation orchestration, enabling specialized agents to collaborate on complex objectives.

Scaling tool use in LMs has since become a central concern.
Toolformer demonstrated that LMs can be trained to make self-supervised API calls~\cite{schick2023toolformer}, while more recent work such as RAG-MCP addresses prompt bloat in tool selection by leveraging retrieval-augmented generation to select only relevant capabilities at inference time~\cite{gan2025ragmcp}.
Beyond single-turn execution, memory and retrieval play an equally important role.
Retrieval-Augmented Generation (RAG) integrates external knowledge into LM reasoning~\cite{lewis2020retrieval}, but remains subject to context window limitations, such as the "Lost in the Middle" phenomenon~\cite{wu2023lost}.
To overcome this, memory-augmented agents like MemGPT~\cite{packer2023memgpt} and subsequent surveys~\cite{wang2024limitssurvey} have highlighted methods for extending long-context reasoning.
Other work formalizes agent memory as a first-class capability, emphasizing the ability to retain past interactions to inform future steps~\cite{sapkota2025ai}.
Recent examples include SciBORG~\cite{muhoberac2025statememoryneedrobust}, a modular agentic framework for scientific tasks that employs a finite-state memory and context awareness to provide reliable execution and interpretable state transitions.
Parallel efforts in task-oriented dialogue also offer complementary insights into grounded task extraction and intent modeling~\cite{qin2023endtoendtask}.

In summary, while prior work has advanced tool use and memory for language-model agents, it generally stops short of what is needed to run a real control room.
These systems usually do not connect directly to control stacks such as EPICS or LabVIEW~\cite{labview2024}, do not detect when a planned action would write to hardware, and do not provide an auditable plan that operators can review before anything is executed.
They also rarely address practical issues for large scale scientific facilities.
Examples include moving large time-series datasets without overloading model context, keeping channel-resolved data in structured form, and running generated analysis code in sandboxed environments with separate read-only and write-enabled access.
In the remainder of this paper (Sections~\ref{sec:task-extraction}–\ref{sec:execution}), we show how Osprey tackles these control-system-specific requirements through plan-first orchestration, safe integration with live control and archiver systems, and deployment patterns that are already used in day-to-day operations at the ALS.

\subsection{Planning and Orchestration}

Planning and orchestration in AI systems trace back to classical approaches such as STRIPS~\cite{fikes1971strips}, PDDL~\cite{mcdermott1998pddl}, and hierarchical task network (HTN) planning~\cite{erol1994htn}.
Recent LM-based approaches build on these ideas by incorporating structured reasoning directly into language models, exemplified by Chain-of-Thought prompting~\cite{wei2023chainofthought} and Pre-Act, which combines planning with explicit reasoning steps~\cite{rawat2025preact}.
DSPy further systematizes this direction, offering a modular framework for program synthesis and prompt optimization that reduces reliance on brittle, handcrafted prompt engineering~\cite{khattab2024dspy,khattab2022demonstrate}.

Complementing symbolic planning, graph-based architectures have emerged as powerful orchestration substrates for agents.
LangGraph provides a task graph engine that enables explicit modeling of execution dependencies~\cite{langgraph2025}, while frameworks like PydanticAI contribute schema validation for structured outputs~\cite{pydanticai2025}.
Related libraries, such as Instructor~\cite{liu2024instructor} and Outlines~\cite{willard2023efficient}, offer constrained generation techniques that increase reliability in tool invocation.

Safety and human oversight remain critical in applying these systems to scientific and industrial domains.
Reviews of self-driving laboratories emphasize that autonomy must work with human oversight for steering and safety~\cite{hysmith2024selfdriving}. Human-in-the-loop Bayesian experiment planning shows how this oversight can be embedded directly into experimental design~\cite{adams2024human}.
More broadly, critical domains such as energy systems, particle accelerators, and laboratories require workflows with check-pointing, explicit approval points and safety boundaries.

Together, these observations motivate the plan-first architecture of our framework, detailed in Section~\ref{sec:orchestration}, where executable plans are generated with explicit dependencies and optional human review before execution.

\subsection{Domain-Specific Applications}

Agentic AI has seen rapid adoption in scientific and industrial applications, where domain-specific requirements demand specialized integrations.
In the natural sciences, autonomous laboratories have been identified as a key driver of accelerated discovery~\cite{Szymanski2023}, supported by robotic labs and LM-driven pipelines for materials exploration~\cite{vriza2025roboticlabs}.
Within imaging sciences, the PEAR framework has demonstrated multi-agent LM automation for ptychography workflows~\cite{yin2024pear}, while other efforts have highlighted emergent autonomous research capabilities of LMs across chemistry and materials science~\cite{ramos2025llmautonomouschem}.
Several domain-focused systems illustrate this diversity: ChemCrow augments LMs with chemistry-specific tools~\cite{bran2024chemcrow}, Co-scientist enables autonomous experimental planning in chemistry~\cite{boiko2023coscientist}, and CRISPR-GPT applies agentic orchestration to gene-editing workflows~\cite{qu2025crisprgpt}.

Among large-scale scientific facilities, agentic approaches have also been explored for particle accelerators and synchrotron beamlines.
GAIA demonstrated an early accelerator operations assistant combining retrieval, scripting, and control~\cite{mayet2024gaia}, while VISION developed a modular AI assistant for beamline experiments, focusing on natural human–instrument interaction~\cite{Mathur_2025}.
These systems validate the potential of agentic AI in facility operations but typically implement custom, facility-specific architectures.
Osprey builds on these pioneering efforts and provides reusable patterns (tutorial implementations, example capabilities, deployment configurations) enabling facilities to adapt the architecture to their specific control environments, as demonstrated through the control assistant tutorial and ALS deployment detailed in Section~\ref{sec:execution}.

\begin{figure*}[t!]
  \centering
  \includegraphics[width=1\linewidth]{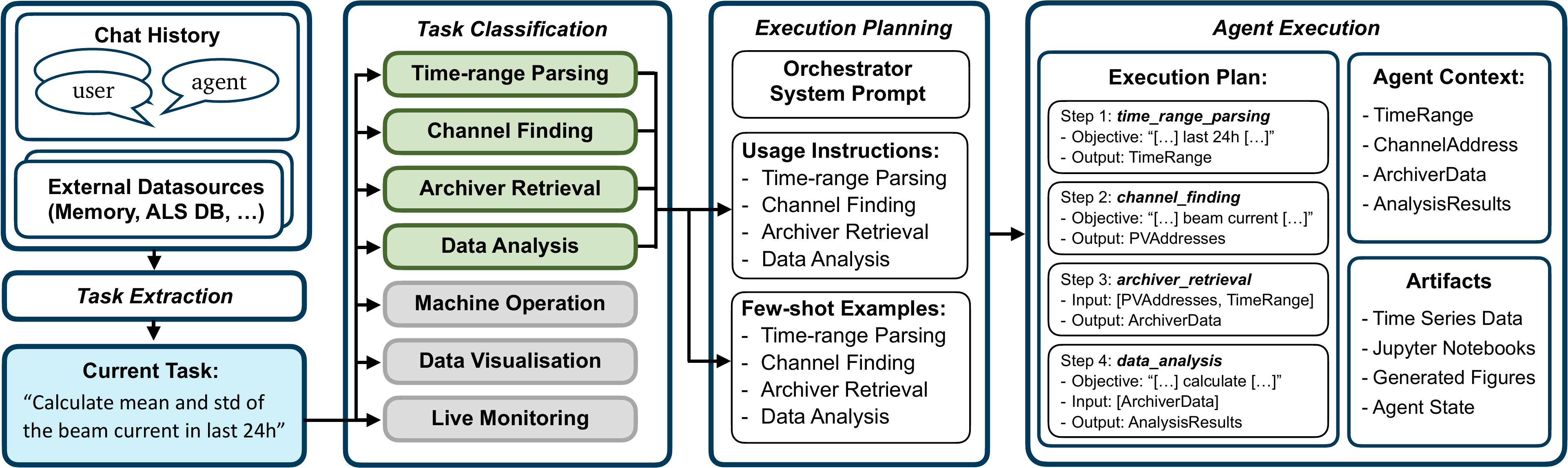}
  \caption{Detailed workflow of the Osprey orchestration layer. Multi-turn conversational inputs and facility-specific data sources (channel databases, archiver systems, operational memory) are transformed into structured task descriptions with resolved control system context. The classifier dynamically identifies relevant capabilities from the available set (channel finding, data retrieval, machine operations, etc.) and passes selected capabilities to the orchestrator. The orchestrator generates a complete execution plan with explicit dependencies and safety annotations, which undergoes pattern detection to identify control system write operations. Plans requiring hardware interaction pause for operator approval before execution. The agent then executes each step with context tracking, artifact management, and containerized isolation. The capabilities illustrated represent the ALS Accelerator Assistant deployment managing hundreds of thousands of control channels; the architecture supports facility-specific capability sets through the registry system.}
  \label{fig:workflow}
\end{figure*}

\section{The Osprey Framework}
\label{sec:framework}

This section explains how Osprey turns natural-language operator requests into safe, executable workflows.
We move from high-level design principles to orchestration, connectors, code generation, and deployment.

Osprey is designed for deploying agentic AI in large-scale, safety-critical control system environments.
Its architecture reflects the core requirements of the aforementioned settings: transparent multi-step planning, robust integration with heterogeneous control stacks, scalable capability management, and isolated execution with explicit safety enforcement.
Natural language requests are transformed into structured, auditable workflows that behave consistently in development, mock-testing, and production control rooms.

Osprey follows a centrally coordinated execution model: a single plan-first orchestrator interprets the operator’s intent and routes tasks through a set of well-defined capabilities.
These capabilities are not autonomous agents acting independently; instead, each contributes a specific, narrowly scoped function to the overall workflow, with all data exchange and decision flow mediated by the orchestrator.
This structure provides clear boundaries, predictable behavior, and transparent coordination.
At the same time, it allows externally defined multi-agent services, such as MCP-based toolchains, to be integrated as a single callable unit within the same planning framework.
This layered architecture mirrors the framework’s operational workflow and aligns with how facilities integrate the system in practice (as illustrated later in the case studies in Section~\ref{sec:controlassistant} and Section~\ref{sec:als}).

\subsection{Design Principles}
\label{sec:design-principles}

Before diving into implementation details, we summarize the main principles that guided the design of Osprey.
These principles capture what safety-critical control rooms need from an agentic system and provide the rationale for the architectural choices in the following subsections.

Osprey’s architecture is guided by several principles motivated directly by control system constraints:

\begin{itemize}
    \item \textbf{Plan-first transparency.}
    The system produces complete execution plans with explicit dependencies
    before invoking any capability. This allows operators to see what the agent
    intends to do, including resolved channels, analysis steps, and any proposed
    hardware writes—before execution.

    \item \textbf{Control-system awareness.}
    Write operations in accelerator and facility control systems are
    safety-critical. Pattern detection and static analysis identify proposed
    control interactions, enforce approval workflows, and verify PV limits using
    facility-provided boundary databases.

    \item \textbf{Protocol agnosticism.}
    Connector abstractions decouple the agent from specific control system
    implementations (such as EPICS gateways), archiver appliances, or mock systems.
    The same workflow operates across different facilities or environments without modification.

    \item \textbf{Scalable capability inventories.}
    A relevance classifier selects only the capabilities needed for the current
    task, preventing prompt inflation as facilities add more services to their
    agentic ecosystem.

    \item \textbf{Isolated, auditable execution.}
    Generated Python code runs in containerized read-only or write-enabled
    environments, and all execution artifacts—such as plans, notebooks, and control-system
    interactions—are recorded for inspection and long-term reproducibility.
\end{itemize}

These principles underpin the layers described below and enable Osprey to function as a production-ready orchestration system, rather than a general-purpose assistant.

\subsection{Orchestration Layer}
\label{sec:orchestration-layer}

Building on these principles, the orchestration layer is where operator intent is first turned into a concrete plan.
It distills conversations into structured tasks, selects the relevant capabilities, and produces a complete execution plan that is executed under safety checks.

This layer forms the interface between natural-language operator instructions and the framework’s structured, multi-step workflows.
It interacts with language models through a model-agnostic interface that supports both remote frontier models and local engines such as Ollama~\cite{ollama} or vLLM~\cite{kwon2023efficient}.
No domain-specific fine-tuning is required; facility behavior is introduced through prompt design, curated example scripts, and the connector and capability abstractions described below.
This keeps deployments fully configurable across facilities while allowing the system to benefit directly from advances in general-purpose model capability.
In practice, deployments vary, but we find that \emph{Claude 4.5 Haiku} offers an excellent balance of performance and low latency, and we recommend it as a practical default.

Within this architecture, the orchestration layer converts natural-language requests into executable workflows in four stages.
It (1) distills the conversation into a structured task, (2) selects only the capabilities relevant to that task, (3) generates a complete, dependency-aware execution plan, and (4) executes the plan under safety checks for all control-system interactions.
This complete workflow is illustrated in \figref{fig:workflow}.

\subsubsection{Task Extraction: From Conversation to Actionable Tasks}
\label{sec:task-extraction}

Extended operator–assistant interactions, especially in facility control rooms, often contain implicitly specified instructions, domain-specific shorthand, and references to earlier steps.
Raw conversation logs are not directly usable for planning: they increase token cost, include details that are irrelevant to execution, and obscure the actionable goal.
Osprey uses LM reasoning to distill this multi-turn context into a structured task containing explicit objectives, constraints, inferred dependencies, and required inputs.

This transformation incorporates multiple knowledge sources such as conversation history, a user-specific operational memory maintained by the framework, or facility-specific knowledge bases, to produce an unambiguous representation of the operator's intent.
The resulting task becomes the foundation for capability selection and execution planning.

\subsubsection{Capability Classification: Adaptive Tool Selection}
\label{sec:classification}

Large facilities expose diverse capabilities for diagnostics, channel finding, archiver access, simulation, or machine operations.
Including the full capability inventory in every prompt would quickly lead to prompt explosion, especially in domains largely absent from model pretraining (e.g., accelerator physics).

Osprey introduces a per-capability binary relevance test: \textit{Is this capability required for the current task?} Few-shot examples associated with each capability guide the LM to make this decision using the structured task as input.
Only the tools classified as relevant should proceed to planning.
This ensures prompts remain compact and well-focused even as capability inventories grow and evolve.

\subsubsection{Planning: Upfront Orchestration}
\label{sec:orchestration}

Reactive agent frameworks provide minimal visibility into future steps.
This is incompatible with safety requirements in accelerator operations: operators must know precisely what PVs will be modified, in what order, and with which inputs; \emph{before} any action occurs.

Osprey therefore adopts a plan-first architecture.
The orchestrator produces a complete execution plan listing all steps, their dependencies, inputs, outputs, and the capabilities involved.
Operators can view, modify, or reject this plan before execution.
Clear dependencies ensure that data flows correctly across steps: for example, semantic channel resolution feeds directly into archiver queries, and retrieved time-series data feeds into Python visualization.

Plan review can be always-on (via configuration) or selectively enabled using runtime commands such as \texttt{/planning:on}.
All interfaces (CLI, web chat, or through APIs) expose the same interrupt-based plan preview.

\subsubsection{Python Code Generation}
\label{sec:code-generation}

Many control-room tasks require the system to generate Python scripts on the fly that follow local conventions for diagnostics, visualization, and machine procedures.
Osprey includes a modular code-generation layer for this purpose.

In this layer, a \emph{generator} is a component that takes a structured task description and produces the corresponding Python code, for example using a language model, a simple template engine, or a deterministic mock backend.
Generators connect to the framework through a lightweight interface, so facilities can choose or implement their own without changing the rest of Osprey.

Three generators are included: a fast single-pass LLM generator, an advanced Claude-based generator using Anthropic’s Agent SDK~\cite{anthropic2025_claude_agent_sdk}, and a deterministic mock generator for testing.
The Claude-based generator supports multi-phase workflows (scan→plan→implement) and can ingest curated example scripts to learn local analysis routines, visualization styles, and device-manipulation procedures.
Facilities configure these workflows entirely through YAML configuration files, choosing between fast profiles for routine use and more robust profiles for complex tasks.

To support facility-specific conventions, the Claude-based generator automatically adopts patterns from these curated examples such as standard visualization styles, diagnostic processing routines, or machine-manipulation procedures used in local operations.
This search operates entirely inside a read-only, sandboxed directory, ensuring that code discovery remains safe and restricted to example paths explicitly designated by the facility.

\subsubsection{Execution: Safety-Enforced and Production-Ready}
\label{sec:execution}

Execution integrates LangGraph checkpointing with control-specific mechanisms for reliability and safety.
Failures such as transient communication errors, invalid channels, or connector timeouts trigger structured retries or re-planning with updated constraints.

At execution time, Osprey enforces a multi-layer, fail-secure safety architecture.
For any step that may interact with the control system, safety enforcement is performed through two complementary analyses:
\begin{itemize}
    \item \textbf{Pattern detection} scans generated Python code for control
    operations such as \texttt{caget}, \texttt{caput}, or vendor-specific write
    calls using configurable regex patterns.
    \item \textbf{PV boundary checking} evaluates proposed setpoints against
    facility-defined operating limits and channel-level access policies, including
    whitelists, read/write flags, and maximum step sizes.
    Requests that violate these constraints or target non-whitelisted channels
    are rejected and require explicit operator approval.
\end{itemize}
Together with code-generation isolation and human-in-the-loop plan and code review,
these checks implement a defense-in-depth safety architecture in which multiple,
independent layers can veto an unsafe operation before it reaches hardware.

By default, code that contains only read operations (or no control-system access at all)
runs automatically in an isolated read-only Jupyter\cite{jupyter2025} container.
Any code that may write to hardware instead requires operator approval and is
presented within a Jupyter notebook review package that highlights the identified
control interactions and PV limit evaluations.
Both policies are configurable: facilities can require review for all generated
code and can choose between containerized execution or a local Python environment,
while still benefiting from the same safety checks.
The use of containers via Docker~\cite{docker_software} or Podman~\cite{podman_github}
separates read-only and write-enabled kernels, protects control networks through
connector-level access control, and ensures that all executions remain auditable
and reproducible.

\begin{figure}[h!]
    \centering
    \includegraphics[width=1\linewidth]{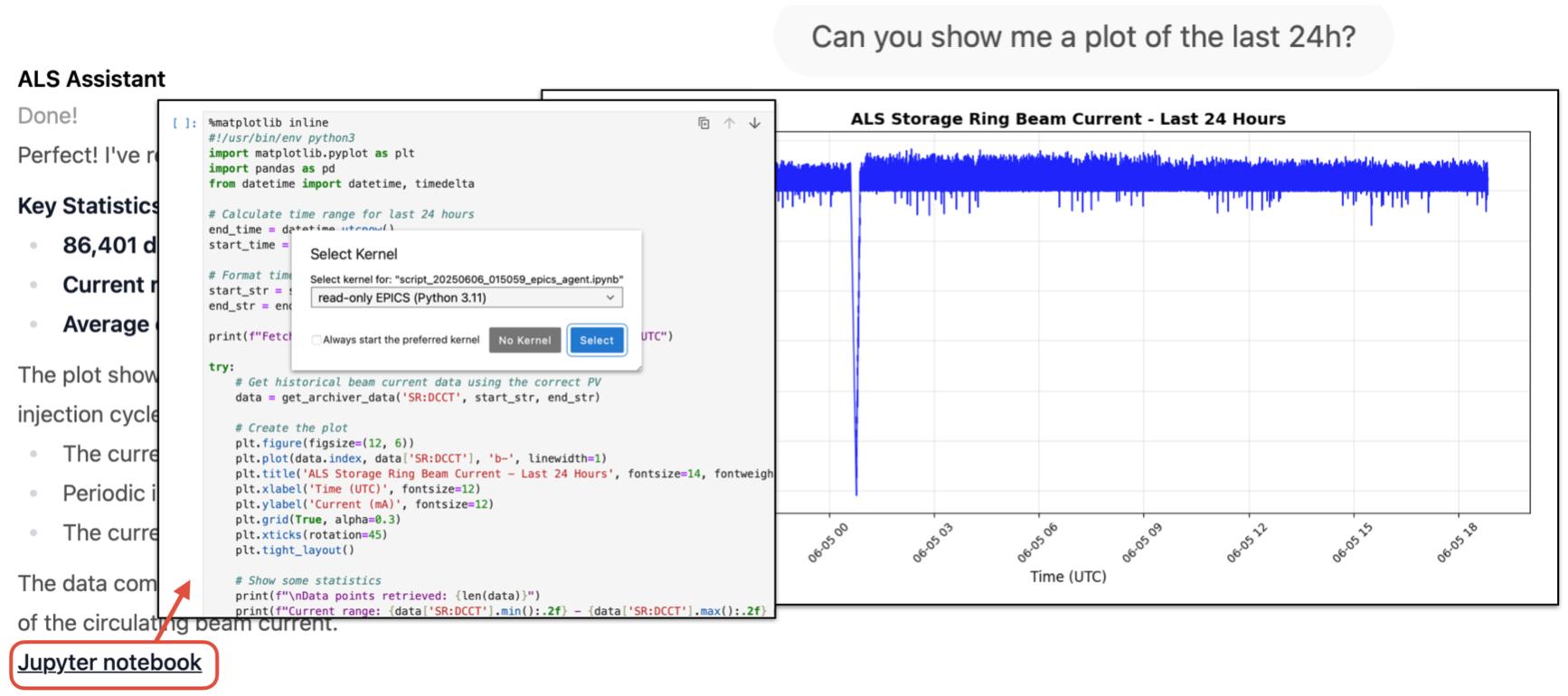}
    \caption{Jupyter Notebook view in OpenWebUI, exposing the Python script that produced the plotted archiver data and allowing users to inspect or rerun any Python-based execution.}
    \label{fig:jupyter}
\end{figure}

\subsection{Facility Integration}
\label{sec:integration}

Different facilities use different control systems, data formats, and operational tools, and Osprey is designed to work with this diversity rather than require standardization.
The framework provides protocol-agnostic connectors that adapt to local infrastructure and a registry system that lets facilities customize components while still receiving framework updates.
This subsection describes how this approach enables integration with existing services and supports deployments from development environments to production control rooms.

\subsubsection{Data and Control Connectors}
\label{sec:connectors}

Reliable control room operation depends on accessing many different subsystems through a single, coherent interface: control-system gateways, archiver appliances, device databases, and mock environments.
For example, facilities using EPICS access process variables through network gateways that expose PVs from the underlying control network, while other control systems provide similar abstractions using different protocols and naming conventions.
Osprey provides connector abstractions that hide these protocol details while exposing a common set of operations for channel read/write, historical data access, and metadata queries.
The same capability definition can run against mock services during development or live control infrastructure in production, with the target selected purely through configuration rather than code changes.

Connectors also encode facility safety policies.
Facilities can configure separate read-only and write-enabled endpoints (for example, distinct EPICS gateways), and Osprey routes operations accordingly.
Read-only diagnostics use the restricted path, while any hardware-writing step that passes the approval workflow in Sec.~\ref{sec:execution} is executed through the write-enabled path.
This adds a network-level safety layer that complements pattern detection and PV boundary checks.

On the data side, connectors return structured objects instead of raw text.
For example, time-series queries from archiver systems return records with values, timestamps, units, alarm states, and operating limits.
These objects can provide compact summaries or downsampled views for language-model reasoning while keeping the full-resolution data available to generated Python code.
This separation between “data for reasoning” and “data for computation” prevents context-window saturation and supports type-safe data propagation across complex workflows.

\subsubsection{Deployment and Operational Readiness}
\label{sec:deployment}

Osprey is deployed as a containerized stack that manages agent orchestration, LLM access, capability services, and isolated Python execution environments.
Its design supports development workstations, control-room servers, and facility clusters via the same declarative configuration model, allowing facilities to move from mock deployments to production control rooms without changing the logical workflow.

Beyond built-in capabilities, Osprey supports automated integration of external tools via the Model Context Protocol (MCP).
Given an MCP server URL, the framework creates a ReAct-style agent whose tools are defined by that server and exposes this agent as a single composite capability.
For the main orchestrator, this MCP-backed capability behaves like any other: it appears in the classification step, can be selected in execution plans, and returns structured results for later steps.
Classifier examples and orchestrator guides for this capability are derived automatically using an LLM, so MCP ecosystems can be brought into the workflow with minimal configuration effort.

Safety and approval checks remain enforced at the Osprey boundary.
Generated Python code and direct interactions with control-system connectors still pass through pattern detection, PV boundary checks, and operator-approval workflows as described in Sec.~\ref{sec:execution}.

Operational transparency is maintained by recording all artifacts produced during a run like execution plans, notebooks, logs, and control-system interactions, and storing them in a form that can be inspected or replayed later.
Execution plans are serialized as structured objects, so operators and developers can review exactly which capabilities were called, in what order, and with which inputs and outputs.
A registry system keeps facility-specific components (such as control-system connector implementations, capability definitions, or custom Python code generators) separate from the shared framework code.
This separation lets facilities customize and extend Osprey for their own infrastructure while still adopting framework updates without maintaining a separate fork.

\section{Case Studies}
\label{sec:casestudies}

To illustrate how Osprey works in practice, we now present two case studies.
The first uses the Control Assistant tutorial in a mock environment with production-inspired patterns\cite{osprey_tutorial_channel_finder}, and the second reports on a production deployment at the Advanced Light Source (ALS) under real accelerator operating conditions, where the framework interacts with hundreds of thousands of control channels and must satisfy facility safety and reliability requirements~\cite{Hellert2024zrj,hellert2025agenticai}.
Together, these case studies show that Osprey is both accessible for initial adoption and robust enough for routine use in a demanding control room.

\subsection{Control Assistant Tutorial}
\label{sec:controlassistant}

The Control Assistant tutorial provides a complete reference implementation of Osprey in a controlled environment, demonstrating how conversational requests are translated into multi-step workflows involving data retrieval, analysis, and visualization.
It includes semantic channel finding, archiver integration, mock connectors for development without hardware access, and containerized Python execution.
Facilities can use this template directly or adapt its patterns to their own control system architectures.\cite{alpha_berkeley_repo,osprey_tutorial_channel_finder}

A representative example is the operator-style prompt:

\begin{quote}
\textit{``Give me a time series and a correlation plot of all horizontal BPM positions over the last 24 hours.''}
\end{quote}

Although concise in natural language, this request requires several coordinated steps typical of real control room workflows.
The system must determine which channels correspond to "horizontal BPM position," interpret the relative time window, retrieve the appropriate historical records from the archiver, and generate Python code to visualize the time series and their correlations.
These tasks normally require facility-specific expertise and manual scripting.

\begin{figure}[t]
    \centering
    \includegraphics[width=0.5\textwidth]{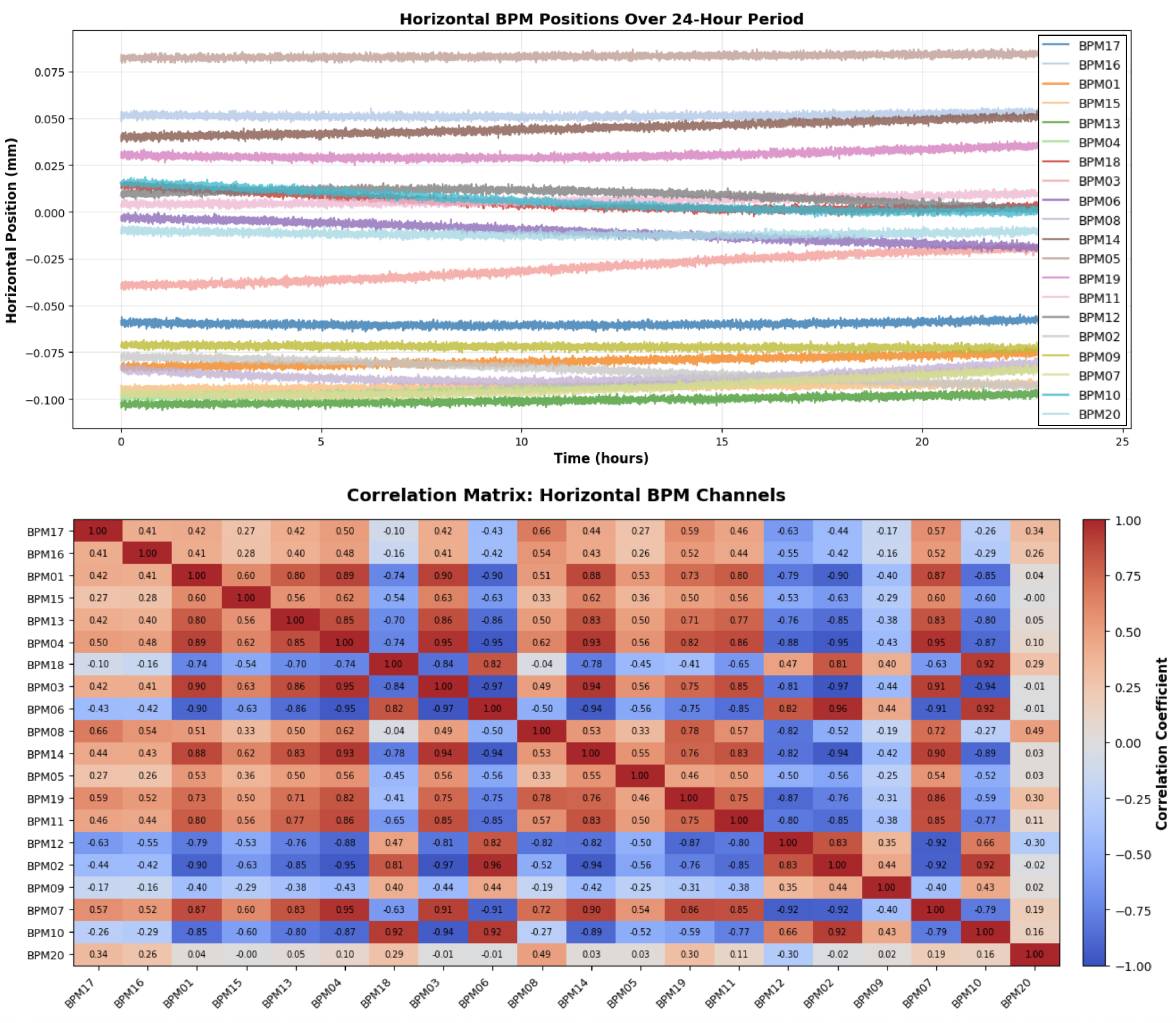}
    \caption{
        Example output from the Control Assistant tutorial for the request
        \textit{``Give me a time series and a correlation plot of all
        horizontal BPM positions over the last 24 hours.''} The framework
        resolves the relevant channels, retrieves historical data via the mock
        archiver, and executes generated Python code in an isolated Jupyter
        environment to produce the plots.}
    \label{fig:control_example}
\end{figure}

Osprey handles this request by breaking it into discrete steps, each addressing a specific sub-task: finding the relevant channel addresses in the control system, parsing the time expression into absolute timestamps, retrieving historical data from the archiver, and generating Python code to create the requested visualizations.
As shown in \figref{fig:plan_bpm}, the framework identifies which operations are independent (such as finding channels and parsing time ranges) and which must wait for earlier results (data retrieval requires both channel addresses and time boundaries).
The resulting plots and execution metadata are returned to the user, as shown in Fig.~\ref{fig:control_example}.

This approach keeps large datasets out of the language model's context by storing time-series data in structured objects that downstream code can access directly, rather than serializing thousands of data points as text tokens.

\begin{figure}
\definecolor{ospreyblue}{RGB}{70,130,180}
\definecolor{capabilitycolor}{RGB}{240,248,255}
\definecolor{headercolor}{RGB}{230,240,250}
\small
\centering
\fbox{
\begin{minipage}{0.95\linewidth}
\vspace{0.3em}

\vspace{0.5em}

\colorbox{headercolor}{
\begin{minipage}{0.98\linewidth}
\vspace{0.2em}
\textbf{Task Extraction} $\rightarrow$ Structured task created \quad | \quad \textbf{Classification} $\rightarrow$ 4 capabilities \quad | \quad \textbf{Planning} $\rightarrow$ 5-step plan
\vspace{0.2em}
\end{minipage}
}

\vspace{0.5em}

\noindent\colorbox{capabilitycolor}{
\begin{minipage}{0.98\linewidth}
\vspace{0.2em}
\textbf{Step 1/5:} \textcolor{ospreyblue}{\textsf{time\_range\_parsing}}
\vspace{0.1em}
\begin{itemize}
  \item[\textcolor{ospreyblue}{$\triangleright$}] \textbf{Input:} \textit{last 24 hours}
  \item[\textcolor{ospreyblue}{$\triangleright$}] \textbf{Output:} ISO time range for the previous 24\,h
\end{itemize}
\vspace{-0.3em}
\end{minipage}
}

\vspace{0.3em}

\noindent\colorbox{capabilitycolor}{
\begin{minipage}{0.98\linewidth}
\vspace{0.2em}
\textbf{Step 2/5:} \textcolor{ospreyblue}{\textsf{channel\_finding}}
\vspace{0.1em}
\begin{itemize}
  \item[\textcolor{ospreyblue}{$\triangleright$}] \textbf{Input:} \textit{all horizontal BPM positions}
  \item[\textcolor{ospreyblue}{$\triangleright$}] \textbf{Output:} List of horizontal BPM position PVs
\end{itemize}
\vspace{-0.3em}
\end{minipage}
}

\vspace{0.3em}

\noindent\colorbox{capabilitycolor}{
\begin{minipage}{0.98\linewidth}
\vspace{0.2em}
\textbf{Step 3/5:} \textcolor{ospreyblue}{\textsf{archiver\_retrieval}}
\vspace{0.1em}
\begin{itemize}
  \item[\textcolor{ospreyblue}{$\triangleright$}] \textbf{Input:} BPM PVs + time range
  \item[\textcolor{ospreyblue}{$\triangleright$}] \textbf{Output:} Time-series records for all BPMs
\end{itemize}
\vspace{-0.3em}
\end{minipage}
}

\vspace{0.3em}

\noindent\colorbox{capabilitycolor}{
\begin{minipage}{0.98\linewidth}
\vspace{0.2em}
\textbf{Step 4/5:} \textcolor{ospreyblue}{\textsf{python}}
\vspace{0.1em}
\begin{itemize}
  \item[\textcolor{ospreyblue}{$\triangleright$}] \textbf{Input:} Historical BPM data
  \item[\textcolor{ospreyblue}{$\triangleright$}] \textbf{Process:} LLM generates analysis plan and Python script
  \item[\textcolor{ospreyblue}{$\triangleright$}] \textbf{Execute:} Containerized code execution (read-only)
  \item[\textcolor{ospreyblue}{$\triangleright$}] \textbf{Output:} Time-series plots and BPM correlation matrix
\end{itemize}
\vspace{-0.3em}
\end{minipage}
}

\vspace{0.3em}

\noindent\colorbox{capabilitycolor}{
\begin{minipage}{0.98\linewidth}
\vspace{0.2em}
\textbf{Step 5/5:} \textcolor{ospreyblue}{\textsf{respond}}
\vspace{0.1em}
\begin{itemize}
  \item[\textcolor{ospreyblue}{$\triangleright$}] \textbf{Input:} links to plots and code + derived statistics
  \item[\textcolor{ospreyblue}{$\triangleright$}] \textbf{Output:} Operator-facing summary, rendered figures and Jupyter notebook link
\end{itemize}
\vspace{-0.3em}
\end{minipage}
}

\vspace{0.3em}
\end{minipage}
}
\caption{Execution plan generated by Osprey for the request
\textit{``Give me a time series and a correlation plot of all horizontal BPM
positions over the last 24 hours.''} The natural language query is first converted
to a structured task through task extraction. After capability classification,
the orchestrator produces a five-step plan to parse the time range, resolve channel addresses,
retrieve historical data, generate correlation plots, and deliver the final
operator-facing response.}
\label{fig:plan_bpm}
\end{figure}

Even though this example uses only mock connectors, the full safety mechanisms are active.
Before any Python code executes, Osprey performs pattern detection to identify potential control-system write operations such as EPICS \texttt{caput} calls.
Because the generated script contains only read-only analysis, it is executed automatically inside an isolated Jupyter container.
Any detected write operations would instead pause execution for operator approval and be routed through the write-enabled environment described in Sec.~\ref{sec:execution}.
The same workflow applies to capabilities derived from MCP servers, and the tutorial can add these automatically by generating capabilities from the MCP tool schemas.

The Control Assistant tutorial demonstrates how the framework handles multi-step workflows: determining task dependencies, managing execution order, and providing safe environments for both development (with mock connectors) and production deployment (with live control systems).
It provides a practical and reproducible on-ramp for facilities exploring agentic workflows prior to engaging with live control systems.
Beyond the tutorial, this pattern is being adapted for production use at multiple accelerator facilities, demonstrating the framework's broader applicability, with deployment results forthcoming.

\subsection{Advanced Light Source: Production Deployment}
\label{sec:als}

The Advanced Light Source (ALS) at Lawrence Berkeley National Laboratory is a third-generation synchrotron facility with a mature EPICS control system and hundreds of thousands of addressable process variables.
Osprey has been deployed as the \textit{ALS Accelerator Assistant}~\cite{hellert2025agenticai}, where it interacts with the production control infrastructure, resolves operator intent across the PV namespace, retrieves historical data from the archiver appliance, and executes approved procedures under facility safety constraints.
This environment provides a stringent validation of the framework's ability to translate high-level natural-language goals into safe, reproducible, multi-stage operational workflows.

A representative example is a complete hysteresis measurement of insertion device (ID) gap versus vertical beam size.
An operator issued the following single natural-language request:

\begin{quote}
\textit{``Get the minimum and maximum value of all ID gap values in the last three days.
Then write a script which moves each ID from maximum to minimum gap and back while measuring the vertical beam size at beamline~3.1.
Sample the gap range with 30 points, wait 5\,s after each new setpoint for the ID to settle, and measure the beam size 5 times at 5\,Hz.
Return a hysteresis plot of beam size versus gap.''}
\end{quote}

The framework breaks this request into coordinated steps: semantic channel resolution identifies the PVs for all insertion devices and the beamline~3.1 diagnostic $\rightarrow$ archiver retrieval fetches three days of historical gap data $\rightarrow$ a data analysis capability processes this time-series to extract minimum and maximum values for each ID $\rightarrow$ the machine operations capability receives these processed limits and generates a bidirectional scan script with the specified dwell times and averaging $\rightarrow$ a final visualization step creates the hysteresis plot.
All proposed hardware writes undergo pattern detection and PV boundary checks, and the complete plan and generated script are presented to the operator for approval prior to execution.

\begin{figure}[t]
    \centering
    \includegraphics[width=1\linewidth]{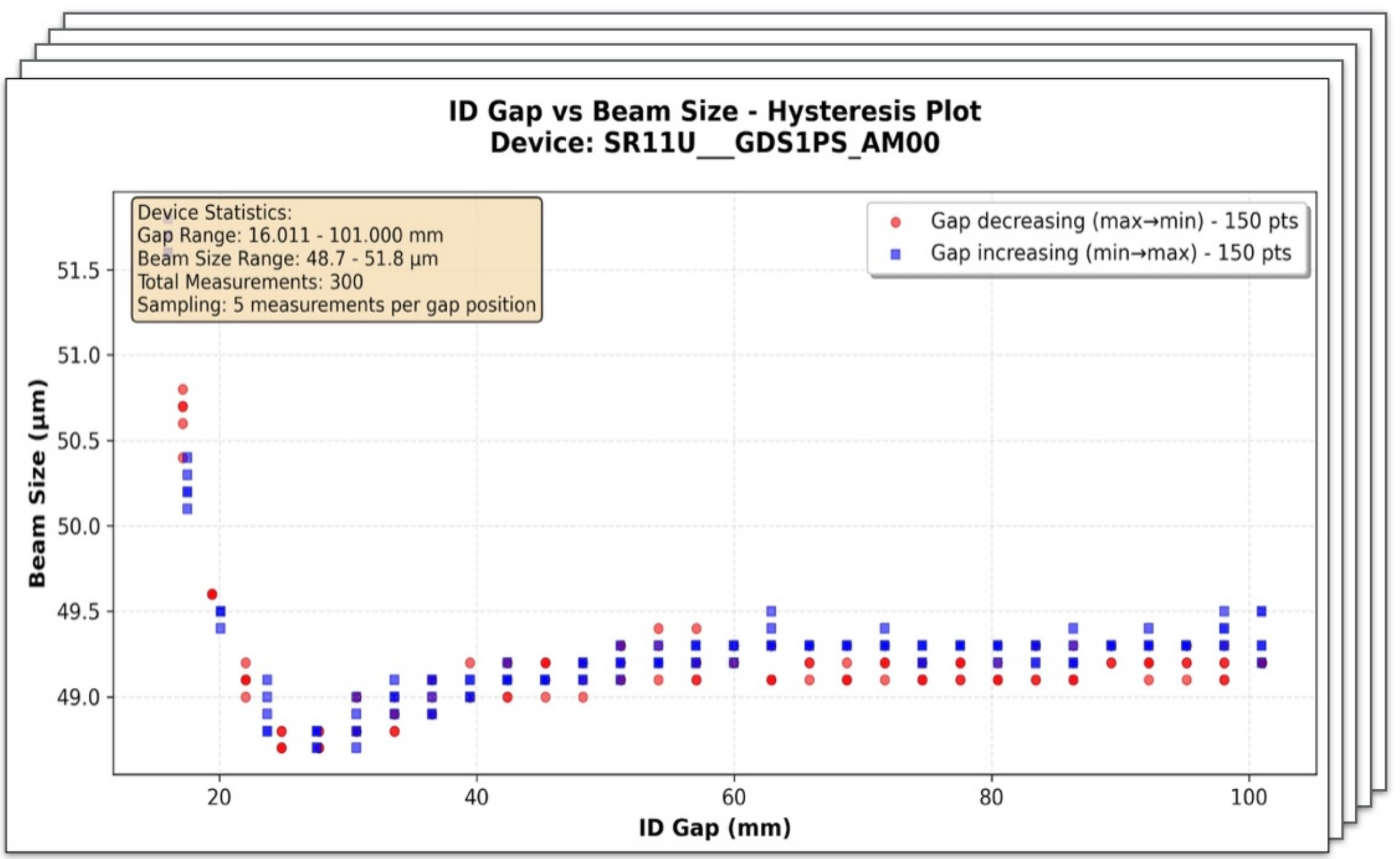}
    \caption{Representative hysteresis measurement of insertion device gap
    versus vertical beam size at the ALS. The execution plan generated by
    Osprey combined historical range extraction, automated script generation,
    and real-time machine control. The agent performed a 30-point
    bidirectional gap sweep with repeated measurements at each point,
    validating safe end-to-end orchestration in a live accelerator
    environment.}
    \label{fig:als_expert_hysteresis_plot}
\end{figure}

Once approved, the measurement runs in a containerized execution environment with controlled access to the EPICS gateways.
The resulting data are processed and shown as a hysteresis curve, shown in Fig.~\ref{fig:als_expert_hysteresis_plot}.
The stability of the vertical beam size across the full range confirms correct machine behavior.
More broadly, this example demonstrates that complex accelerator procedures can be executed from natural language while preserving transparency, operator oversight, and facility safety requirements.

\section{Conclusion}
\label{sec:conclusion}

We have presented Osprey, a framework designed for deploying agentic AI in large, safety-critical control systems.
The architecture reflects the practical constraints of these settings: semantic addressing across large channel namespaces, protocol integration with diverse control systems (such as EPICS and related stacks), strict transparency requirements around planned actions, and isolated execution of generated code.
Plan-first orchestration produces reviewable execution plans before any capability is invoked; capability classification keeps prompts focused as facilities add more tools; connector abstractions provide protocol-agnostic access to control systems, archiver appliances, and mock services; and containerized code execution, combined with pattern detection and PV boundary checks, enforces safety for all hardware-writing operations.

The framework has been deployed in both tutorial and production settings.
The Control Assistant tutorial offers a complete, reproducible template that demonstrates semantic channel finding, archiver integration, and safe Python execution with mock connectors.
At the Advanced Light Source, the \emph{ALS Accelerator Assistant} runs on the live control system, resolving operator queries across hundreds of thousands of PVs and executing multi-step procedures such as insertion-device hysteresis measurements under facility safety constraints.
These case studies indicate that agentic workflows can be introduced into real control rooms while preserving operator oversight, reproducibility, and existing safety processes.

Osprey’s control-system-focused design, combined with automated capability generation from MCP servers, makes it a reusable orchestration layer for scientific facilities that need to combine AI agents with heterogeneous institutional tools.
While our examples focus on particle accelerators, the same patterns—plan-first execution, capability classification, connector-based integration, and safety-aware execution—are applicable to other domains where complex technical systems must be controlled in a transparent and auditable way.
Future work will focus on broader cross-facility deployments, richer validation metrics for agent-assisted operations, and deeper integration with facility data standards and knowledge bases.
\section*{Acknowledgments}

This research used the CBorg AI platform and resources provided by the IT Division at Lawrence Berkeley National Laboratory.
We gratefully acknowledge Andrew Schmeder for his consistent responsiveness and support, which ensured that CBorg served as an invaluable resource for the development of this framework and the ALS expert agent.

We are grateful to Alex Hexemer for his steady encouragement and advocacy in advancing agentic AI as a facility-wide capability at the ALS, which strongly influenced the broader direction of this framework.

We also thank Frank Mayet for his collaborative spirit and for sharing insights from his pioneering Gaia prototype, which helped guide the early trajectory of agentic AI efforts at the ALS.

We warmly thank our colleagues at the ALS, Fernando Sannibale, Marco Venturini, Simon Leemann, Drew Bertwistle, Erik Wallen, Hiroshi Nishimura, Tom Scarvie, Christoph Steier, Tynan Ford, and Edison Lam, for their expertise, feedback, and support in exploring and shaping the integration of Osprey into accelerator control operations.

We further acknowledge the use of AI tools during the preparation of this work.
Cursor, primarily with Claude 4, was employed extensively during development, while GPT-5 was used to refine the language of this manuscript.

This work was supported by the Director of the Office of Science of the U.S.
Department of Energy under Contract No. DE-AC02-05CH11231.

\bibliographystyle{apsrev4-2} \bibliography{references}

\end{document}